\documentclass[11pt,twoside]{article}
\usepackage{FUSE2004}
\usepackage{natbib}

\usepackage{epsf}
\usepackage{psfig}
\usepackage{lscape}

\begin{document}
\title{FUSE Observations of Active Binaries}
\author{Cynthia S. Froning}
\affil{Center for Astrophysics and Space Astronomy, University of Colorado,
593 UCB, Boulder, CO 80309, USA}

\begin{abstract}

FUSE observations have been instrumental in advancing our understanding of
the physical properties and behavior of active binary systems, including
cataclysmic variables (CVs) and X-ray binaries (XRBs). FUSE data have
allowed observers to study: accretion disks, magnetically-channeled
accretion flow, and white dwarf accretors, and how these respond to
accretion fluctuations and disk outbursts; the role of binary evolution in
determining system properties; the vertical and azimuthal structure in
disks and disk winds; and what the variations in active binary properties
reveal about accretion physics in compact systems.  Results of FUSE
observations of active binaries are reviewed here.

\end{abstract}

\section{Introduction}

Active binaries are interacting binary systems ($P_{orb} \simeq 1$ --
20~hr) in which a donor star transfers mass to a compact object.  In
cataclysmic variables (CVs), the compact object is a white dwarf (WD),
while in X-ray binaries (XRBs), the compact object is a neutron star or a
stellar-mass black hole. In CVs and low-mass XRBs, mass is transferred via
Roche lobe overflow of the donor star and in most cases is accreted onto
the compact object through an accretion disk. Active binaries are excellent
test beds for the study of binary evolution, SN Type Ia progenitors, and
accretion and relativistic physics in nearby, unembedded objects.  In the
FUV, active binaries show emission from the WD, the channeled accretion
flow (in magnetic systems), the inner accretion disk ($\leq20 R_{WD}$), and
accretion disk winds. The dominant FUV sources and the system morphologies
and behaviors vary dramatically in active binaries as a function of compact
object type, orbital period, evolutionary history, mass accretion rate, and
viewing inclination.  To date, more than 90 active binaries have been
observed by FUSE: 24 dwarf novae, 25 non-magnetic novalikes, 12
intermediate polars, 17 polars, 4 super-soft XRBs, 5 high-mass XRBs, 2
low-mass XRBs, and 3 unclassified systems.  Below, we summarize some of the
scientific results obtained from these observations and discuss future
work.

\section{Results from FUSE Active Binary Observations}

\subsection{White Dwarfs in CVs}

In quiescent dwarf novae (DN), the accretion disk is in a low state and the
dominant FUV emission source is the WD. The structures and morphologies of
the WDs in CVs are affected by their histories of ongoing accretion and
accretion instabilities.  DN outbursts typically recur on time scales of
days to years.  During outburst, the WD is heated by the accreted material,
followed by cooling during quiescence.  The WD cooling is shown in the FUV
spectra of two DN in Figure~1.  The apparent temperature of the WD in U~Gem
(upper panel) decreases from 43,000~K immediately after outburst to
30,000~K four months into quiescence \citep{froning2001,long2004}.  The
temperature of the WD in VW~Hyi (lower panel) drops from 23,000~K to
18,000~K \citep{godon2004}.

In contrast to field WDs, the spectra of CV WDs show metal absorption
lines, reflecting the composition of the material being accreted onto its
surface.  Detailed models of the metal-enriched WD spectra often show
abundance anomalies.  In MV~Lyr, the WD spectrum can be well fit assuming
0.3~$Z_{\odot}$ metallicity for all species \citep{hoard2004a}, but more
typical are abundance patterns in which C is underabundant and/or
N overabundant, as seen in SS~Aur, EY~Cyg, U~Gem, and WZ~Sge
\citep{sion2004b,sion2004a,froning2001,long2003}.  The abundances anomalies
represent evolutionary effects on the present composition of the systems, a
point that will be discussed in more detail in the next section.

In many, perhaps most, quiescent DN, it is clear that a second source
contributes to the FUV spectrum.  In U~Gem, better model fits and
correspondance between observed FUV flux and temperature declines can be
obtained if it is assumed that 15--20\% of the WD cools from 70,000~K to
30,000~K after outburst rather than the entire WD cooling \citep{long2004}.
Two-temperature fits to the WD spectrum are also used to model the FUV
spectra of SS~Aur, Z~Cam, EY~Cyg, and VW~Hyi
\citep{sion2004b,hartley2004,sion2004a,godon2004}.  The second source may
not be part of the WD, however. In VW~Hyi, variations of up to 20\% are
observed in the FUV spectrum at the shortest wavelengths where the WD does
not contribute \citep{godon2004}. In WZ~Sge, the best fits to the
quiescent spectrum (seen in the bottom panel of Figure~3) are obtained when
the WD atmosphere is veiled by a solar composition absorbing slab of
material along the line of sight \citep{long2003}. The nature of the second
source remains poorly understood, although its behavior is clearly tied to
the outburst cycle.  Possibilities include an accretion belt on the WD
surface, emission from the boundary layer between the disk and the WD, or
residual accretion during quiescence, either from the disk itself or
through coronal processes.

\begin{figure}[!ht]
\plotone{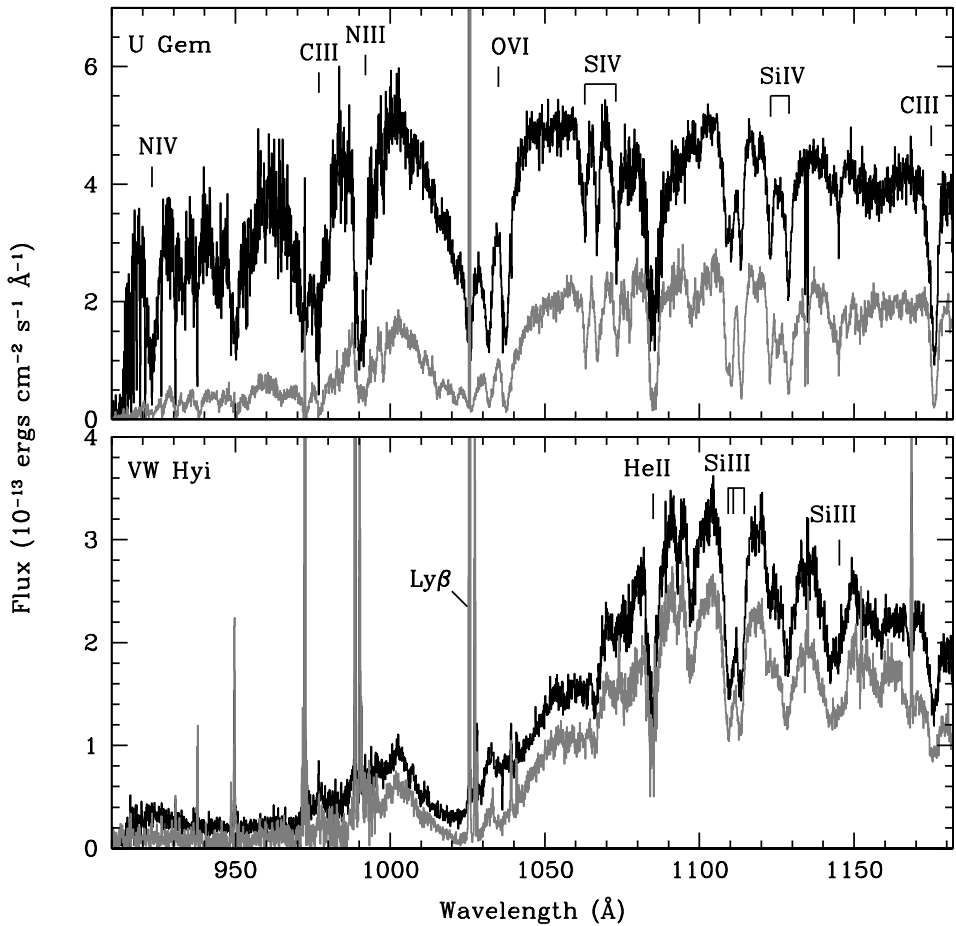} 
\caption{FUSE observations of the DN U~Gem and VW~Hyi in quiescence.  The
upper panel shows the FUV spectrum of U~Gem obtained at two times after
(different) outbursts.  In black is the spectrum obtained immediately after
an outburst, while in gray is a spectrum obtained approximately 4 months
after outburst.  The lower panel shows spectra of VW~Hyi: the spectrum in
black was obtained 11 days after outburst while the spectrum in gray was
obtained about 2 months after the previous outburst.  Prominent lines are
labeled.  All spectra are binned to 0.1~\AA\ dispersion.}
\end{figure}

\subsection{Channeled Accretion Flow in CVs}

For CVs in which the WD accretor has a strong magentic field, the accretion
disk is disrupted and accretion onto the WD occurs via channeled flow along
the magnetic field lines.  For magnetic field strengths $\geq$100~kG, the
disk is only partially disrupted; these are the intermediate polar (IP)
systems.  When the magnetic field strength reaches $B\simeq 10-250$~MG in
the polars, the accretion disk is completely disrupted and material from
the accretion stream is directly entrained onto the WD field lines.  The
FUV emission in these CVs is dominated by the channeled accretion flow and
the accretion poles where the flow impacts the WD surface.  Example FUV
spectra of two magnetic CVs, the IP TW~Pic and the polar AN~UMa, are shown
in Figure~2.  Both are characterized by blue continua on which is
superimposed strong, broad emission lines of CIII, NIII, SiIII, SiIV, SIV,
and OVI.

\begin{figure}[!ht]
\plotone{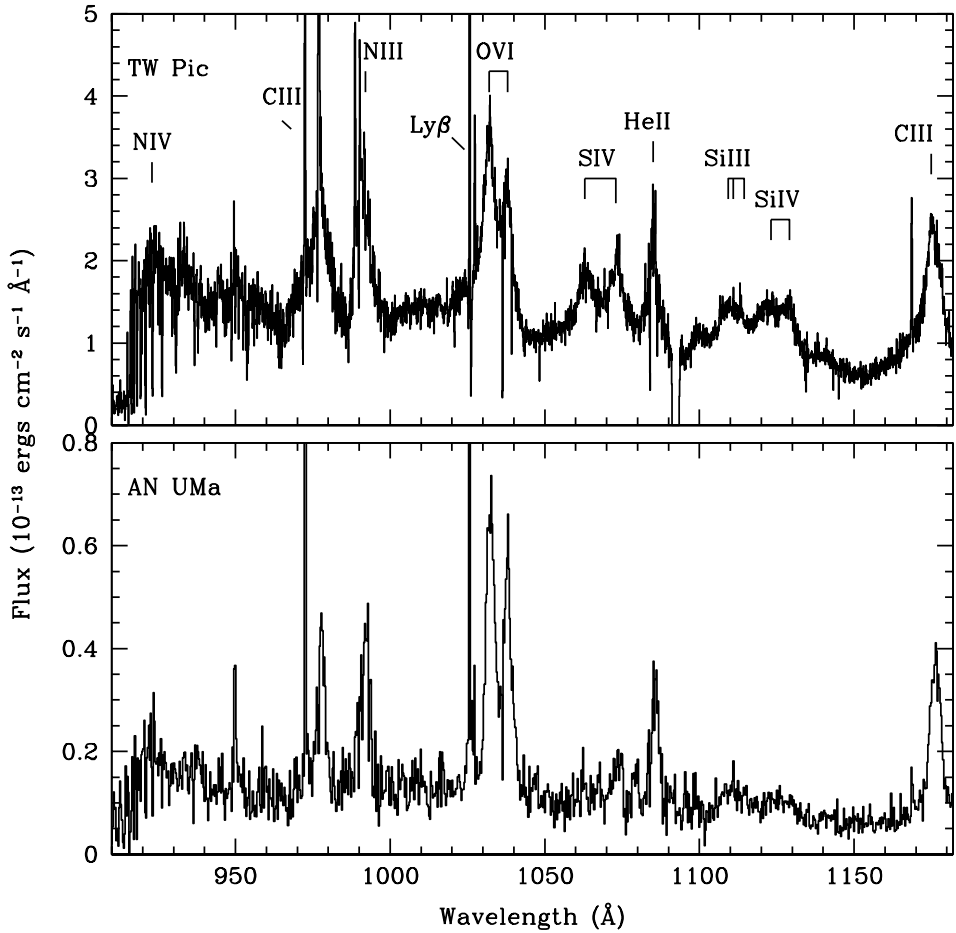}
\caption{Magnetic CVs AN~UMa and TW~Pic.  The top panel shows the spectrum
of the intermediate polar TW~Pic while the bottom panel shows the
time-averaged spectrum of the polar AN~UMa. The TW~Pic spectrum is binned
to 0.1~\AA\ and the spectrum of AN~UMa is binned to 0.3~\AA.}
\end{figure}

To date, analyses of the FUSE observations of the polars AM~Her
\citep{hutchings2002}, BY~Cam \citep{mouchet2003}, VV~Pup
\citep{hoard2002}, and AR~UMa \citep{hoard2004b}; and the IP V405~Aur
\citep{sing2004} have been published. In AM~Her and VV~Pup, the FUV
spectrum is dominated by OVI in emission from the channeled accretion flow
and by continuum emission from the accretion pole ($T \geq 90,000$~K for a
blackbody source in VV~Pup).  The continuum flux increases in both systems
as the dominant accretion pole rotates into the direct line of sight.
V405~Aur also has OVI emission from the accretion flow as well as a broader
emission feature from the accretion disk.  By~Cam shows strong emission
features but with anomalous line ratios: weak C and O lines combined with
strong N emission.  Mouchet et al. demonstrated that the observed line
ratios cannot be reproduced by varying the photoionization structure of the
channeled accretion stream and instead require, as with the WD spectra in
the previous section, non-solar abundances in the accreted material.  These
abundance anomalies provide clues to the role binary evolution plays in
determining the properties of active binaries, but the cause is not yet
known; current models invoke CNO processing in the binary or partial
evolution of the donor star.  FUSE observations have also probed magnetic
field effects on WD atmospheres.  AR~UMa has the strongest magnetic field
strength of any known polar ($B\simeq240$~MG).  FUSE observations taken
when AR~UMa was in a low accretion state show a 20,000~K WD spectrum
modified by the strong magentic field, which manifests itself in Zeeman
absorption features of the Lyman series lines.  Also observed are normally
forbidden Zeeman transitions that become enabled in the presence of the
strong electric fields.

\subsection{Accretion Disks and Outflows in CVs and XRBs}

In high accretion rate CVs, where the disk is more luminous than the WD,
and in XRBs (where there is no WD) the accretion disk is the dominant
source of emission in the FUV.  The FUV continua of active binaries are
sensitive to the mass accretion rate in the disk as the peak temperatures
in the inner disks ($\simeq50,000$~K) leads to a turnover in their spectral
energy distributions in the FUSE waveband.  FUSE spectra have been modeled
with steady-state accretion disk models to measure accretion rates during
high states and on decline from DN outburts
\citep{froning2001,long2003}. An illustration of how the spectra of
quiescent and disk-dominated active binaries differ is given in Figure~3,
which shows how the FUV spectrum of the DN WZ~Sge evolved after its 2001
outburst \citep{long2003}.  At outburst peak (the uppermost panel), the
spectrum is dominated by the accretion disk and a strong disk outflow,
manifested primarily in OVI. The lower three panels show the changes in the
spectrum on the decline from outburst as the wind ceases, the disk fades,
and the cool (23,000~K) WD begins to dominate the FUV spectrum.

\begin{figure}[!ht]
\plotone{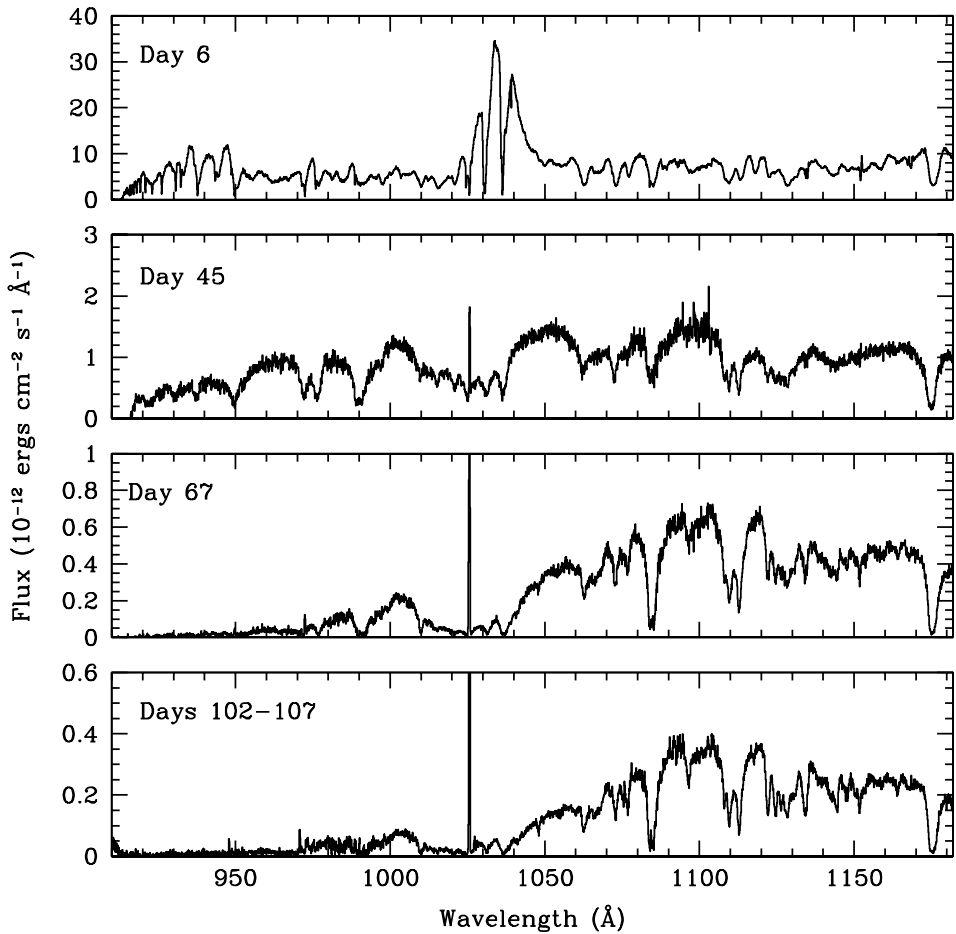}
\caption{The evolution of the FUV spectrum of the DN, WZ~Sge, after
outburst. Figure adapted from Long et al. 2003.}
\end{figure}

FUSE observations have revealed the importance of vertical structure in
accretion disks on the observed behavior of active binaries: simple models
of thin, flat disks are poor descriptions of FUV disk morphologies.  The
spectrum of the DN U~Gem in outburst shows a rich line absorption spectrum
(the spectrum can be seen in the lower panel of Figure~5; Froning et
al. 2001).  The absorption becomes stronger from orbital phases 0.5 -- 0.8,
the same phases at which X-ray and EUV light curve dips are observed in the
system.  The dips and the increased absorption occur at the phases in which
the mass accretion stream from the donor star rotates into the line of
sight and are believed to be caused by a bulge where the stream impacts the
accretion disk or by overflow of the stream above the disk downstream of
the impact point.  Similar azimuthally-dependent increases in line
absorption have been seen in FUSE observations of the DN Z~Cam
\citep{hartley2004}.

Figure~4 shows the effects of azimuthal asymmetries in the disk structure
on the FUV light curve of the eclipsing novalike CV, UX~UMa
\citep{froning2003}.  The upper three panels show the 1~sec FUV light curve
of UX~UMa, constructed from FUSE time-tag event files of the observation.
The light curves exhibit the rapid variability, or flickering, associated
with disks in active binaries.  The bottom panel shows the orbital phase
binned average of the observation.  In addition to the eclipse of the disk
at orbital phase 1, there is a broad dip in the FUV flux centered around
phase 0.6, suggestive of azimuthal changes in the vertical scale height of
the absorbing outer disk. Similar behavior is seen in the super-soft XRB,
RXJ0513--69 \citep{hutchings2002}. In extreme cases, flaring of the outer
disk rim can completely obstruct the view of the inner accretion disk, as
is the case in the high inclination CV, DW~UMa \citep{hoard2003}.  DW~UMa
illustrates the complexity of behavior of active binaries in the FUV, as it
contains both a self-occulting disk and vertically-extended emitting
material that produces a strong emission line spectrum.

\begin{figure}[!ht]
\plotone{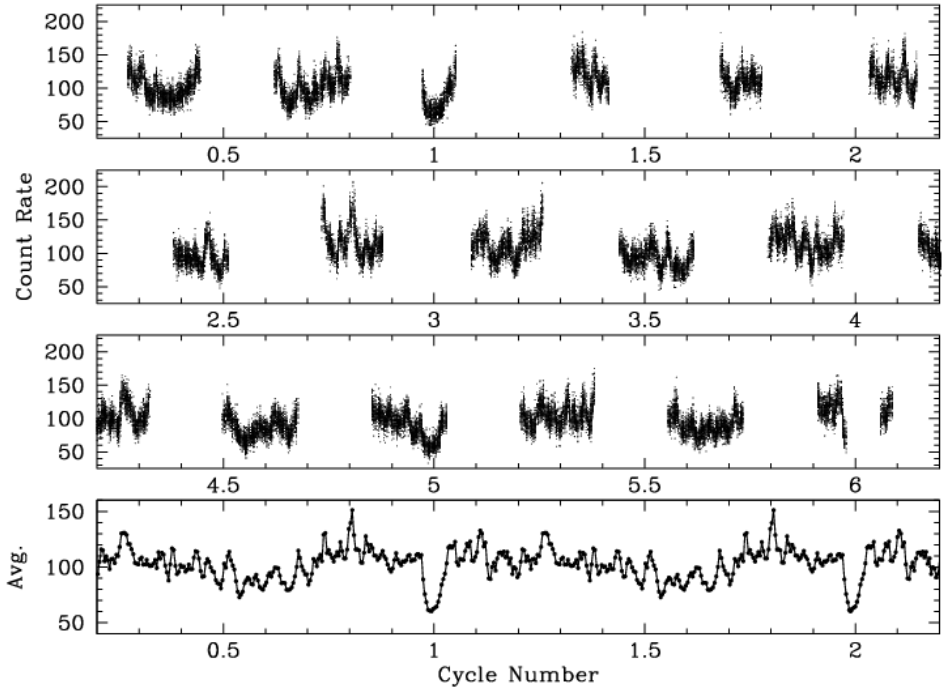}
\caption{High time resolution light curve of UX~UMa.  The top three panels
show the 1~sec average FUV count rate with time from the 2001 observation
of UX~UMa.  The bottom panel shows the orbital phase-binned light curve of
the data. Figure from Froning et al. 2003.}
\end{figure}

The FUV line spectra of disk-dominated active binaries vary widely.  In
many high accretion rate systems, the line emission originates in a disk
outflow in the form of a fast wind.  An example of a wind spectrum is given
in the top panel of Figure~5, which shows the FUV spectrum of the novalike
IX~Vel.  The lines are broad and blue-shifted and appear in a wide range of
species and ionization levels.  As the viewing inclination increases, the
winds appear as P~Cygni profiles and then purely in emission from photons
scattered into the line of sight (as in the spectrum of WZ~Sge in the top
panel of Figure~3).  FUSE observations have shown that the winds in active
binaries are also affected by departures from axisymmetry in the accretion
disk, causing the wind lines to be modulated in strength and profile on the
orbital period \citep{hutchings2001,prinja2003,prinja2004}.  Studies of the
changes in the FUV spectrum of UX~UMa through eclipse have shown that its
wind varies with scale height above the disk, changing from a low outflow
velocity, dense region to the fast, vertically-extended outflow
\citep{froning2003}.

Disk-dominated active binaries do not always show wind lines in the FUV,
however. In the aforementioned outburst spectrum of U~Gem, shown in the
lower panel of Figure~5, the absorption lines are too narrow and at too low
a velocity to originate in a wind.  Aside from the weak signatures of the
wind in OVI, the wind in U~Gem does not appear in the FUV, though a highly
ionized outflow does manifest itself in the EUV.  The super-soft XRB
RXJ0513--69 shows very broad ($\simeq$4000~km~s$^{-1}$) emission lines of
OVI that are believed to originate in the inner accretion disk
\citep{hutchings2002}.  Similarly, the FUV spectra of the super-soft XRB
Cal~83 and the black hole XRB Cyg X-3 show broad OVI emission
\citep{schmidtke2004,hutchings2003}. Emission features in OVI and other
transitions regularly appear in the FUV spectra of CVs in which the mass
accretion rate is too low to sustain a wind.  The emission lines are
believed to originate in a chromosphere or corona overlying the bulk of the
accretion disk, but no models yet exist that fully explain their formation
and the properties of the line emitting region.

\begin{figure}[!ht]
\plotone{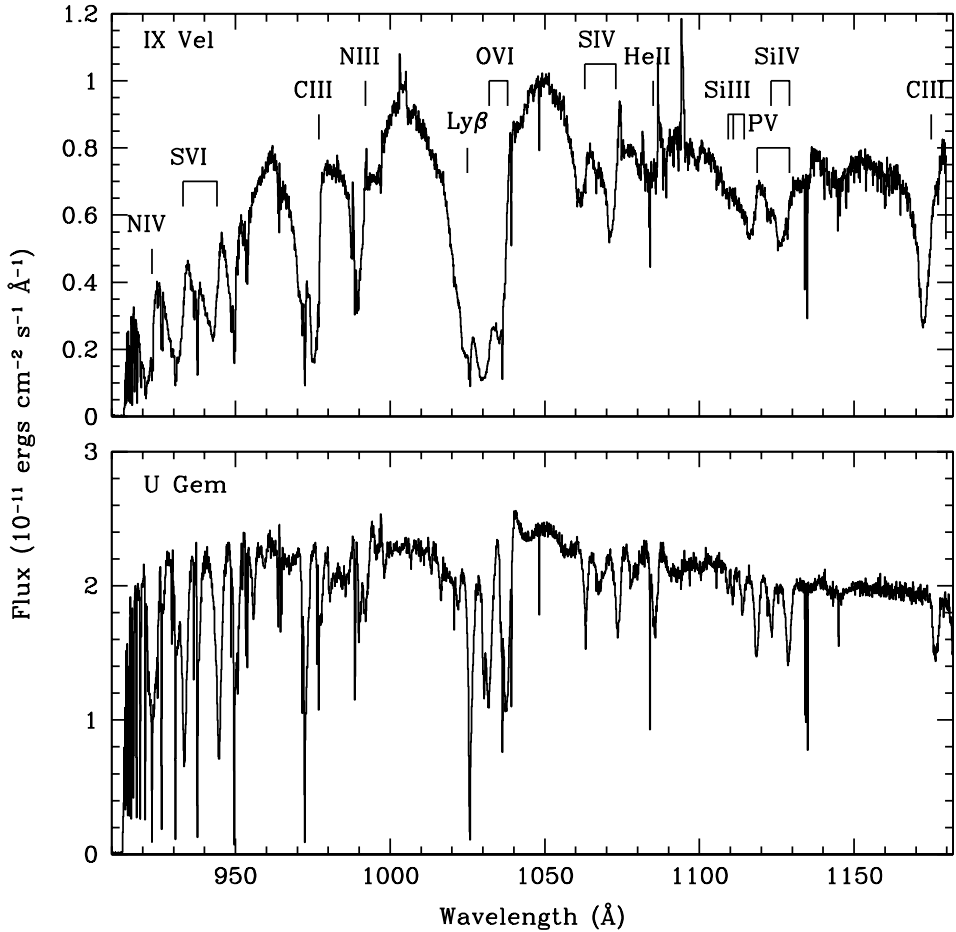}
\caption{Differing behavior in two high accretion rate CVs.  The upper
panel shows the time-averaged spectrum of the NL, IX~Vel.  The lower panel
shows the time-averaged spectrum of the DN, U~Gem, at outburst peak.}
\end{figure}

\section{Conclusions and Future Work}

After five years, FUSE has observed most of the bright CVs and all of the
bright XRBs available to the telescope.  In doing so, it has revealed much
about the properties and behavior of active binaries: the response of disks
and WDs to accretion events; the role of binary evolution in setting system
abundances; the importance of vertical structure in disks and disk
outflows; and the strengths and limitations in our picture of the launching
mechanism for disk winds.  Future work will concentrate on utilizing the
large FUSE database to probe the general properties of active binaries and
on expanding our understanding of accretion physics through intensive
studies and modeling of single systems.  A survey of the FUV properties of
CVs as a function of mass accretion rate, orbital period, evolutionary
history, and binary viewing inclination is ongoing \citep{froning2003}.
Also ongoing are large programs devoted to observations of single systems
that will undertake detailed analysis and modeling of time variability to
determine the structure and properties of the WDs, disk and magnetic
accretion regions, and outflows in active binaries.  FUSE has opened a new
window into the study of active binaries and will continue to provide
cutting-edge scientific observations of these systems for some time to
come.

\end{document}